\documentclass[12pt]{iopart}
\newcommand{\bk}{{\bf k}}

\newcommand{\cH}{{\mathcal H}}
\newcommand{\cZ}{{\mathcal Z}}

\begin{document}
\title[Thermodynamic limit in FES]{The thermodynamic limit for fractional exclusion statistics}
\author{Drago\c s-Victor Anghel}
\address{Department of Theoretical Physics, National Institute for Physics and Nuclear Engineering--''Horia Hulubei'', Str. Atomistilor no.407, P.O.BOX MG-6, Bucharest - Magurele, Romania}
\ead{dragos@theory.nipne.ro}
\begin{abstract}
I discuss Haldane's concept of generalised exclusion statistics 
(Phys. Rev. Lett. {\bf 67}, 937, 1991) and I show that it leads to 
inconsistencies in the calculation of the particle distribution that 
maximizes the partition function. These inconsistencies appear when mutual 
exclusion statistics is manifested between different subspecies of particles 
in the system. In order to eliminate these inconsistencies, 
I introduce new mutual exclusion statistics parameters, which are 
proportional to the dimension of the Hilbert sub-space on which 
they act. These new definitions lead to properly defined particle 
distributions and thermodynamic properties. 
In another paper (arXiv:0710.0728) I show that 
fractional exclusion statistics manifested in general systems with 
interaction have these, physically consistent, statistics parameters.
\end{abstract}
\maketitle

\section{Introduction}

Haldane's concept of fractional exclusion statistics (FES) 
\cite{PhysRevLett.67.937.1991.Haldane} have been applied to the study 
of many types of physical systems, revealing interesting properties. 
For example it has been applied to strongly interacting 
systems, such as Tomonaga-Luttinger model \cite{ProgrTheorPhys.5.544.1950.Tomonaga,JMathPhys.4.1154.1963.Luttinger,JMathPhys.6.304.1965.Mattis,PhysRevLett.81.489.1998.Carmelo}, Colagero-Sutherland model \cite{JMathPhys.10.2191.1969.Colagero,JMathPhys.12.247.1971.Sutherland,PhysRevA.4.2019.1971.Sutherland,PhysRevA.5.1372.1972.Sutherland,PhysRevB.60.6517.1999.Murthy}, fractional quantum Hall effect \cite{PhysRevLett.72.600.1994.Veigy,NuclPhysB470.291.1996.Hansson,IntJModPhysA12.1895.1997.Isakov}, or to interacting particles in one or two-dimensional systems, described in the mean field approximation \cite{PhysRevLett.73.3331.1994.Murthy,PhysRevLett.74.3912.1995.Sen,JPhysB33.3895.2000.Bhaduri,PhysRevLett.86.2930.2001.Hansson}. 
The statistical properties of FES systems 
have been calculated mainly by Isakov \cite{PhysRevLett.73.2150.1994.Isakov} 
and Wu \cite{PhysRevLett.73.922.1994.Wu}, while Iguchi extended the 
Fermi liquid model to the model of a FES liquid 
\cite{PhysRevLett.80.1698.1998.Iguchi,PhysRevB.61.12757.2000.Iguchi}; 
the microscopic reason for the manifestation of FES have also 
been discussed by several authors 
\cite{PhysRevLett.73.3331.1994.Murthy,PhysRevLett.74.3912.1995.Sen,PhysRevB.60.6517.1999.Murthy,NuclPhysB470.291.1996.Hansson,IntJModPhysA12.1895.1997.Isakov,PhysRevLett.86.2930.2001.Hansson,PhysRevLett.85.2781.2000.Iguchi}. 

Although the concept received so much attention and has been applied 
in general to many types of systems, I will show here 
that when mutual exclusion statistics is manifested between different 
subspecies of particles in the system, FES leads to thermodynamic 
inconsistencies. 
I will also show that these inconsistencies can be corrected by a 
redefinition of the exclusion statistics parameters. 

In a related paper I showed that fractional exclusion 
statistics appears in general systems of interacting particles and 
the statistics parameters indeed obey the rules conjectured 
here \cite{submitted.FESinteraction}. 

\section{Thermodynamic inconsistencies in FES}\label{inconsistent}

In this section I will prove using two model systems, that in FES 
systems the equilibrium particle populations are ambiguously defined, 
if \textit{mutual} statistics parameters are not zero. For this, I will 
recalculate the partition function and the 
most probable particle distribution in a FES system, following the 
procedure used by Wu in Ref. \cite{PhysRevLett.73.922.1994.Wu}. 

Haldane defined the fractional exclusion statistics as acting 
on Hilbert spaces of finite dimensions \cite{PhysRevLett.67.937.1991.Haldane}. 
%like the Hilbert space of quasiparticles in a finite region of condensed matter. 
If we have only one such a space, in which we put $N$ ideal bosons or 
fermions, then the number of microscopic configurations we have in 
the system is $W_b=(G+N-1)!/[N!(G-1)!]$ (for bosons) or $W_f=G!/[N!(G-N)!]$ 
(for fermions). Fractional exclusion statistics of parameter $\alpha$ is an 
interpolation between these two cases and the number of configurations 
is $W=[G+(N-1)(1-\alpha)]!/\{N![G-\alpha N-(1-\alpha)]\}$--we say that 
the addition of $\delta N$ particles in the system reduces 
the number of available states in the system by $\alpha\delta N$ 
\cite{PhysRevLett.67.937.1991.Haldane,PhysRevLett.73.922.1994.Wu}. 

Now let us generalize the problem to the case when we have more than 
one Hilbert space. The spaces are $\cH_0$, $\cH_1$, 
\ldots, of dimensions  $G_0$, $G_1$, \ldots, and which contain $N_0$, 
$N_1$, \ldots, particles. In this case we have the FES 
parameters $\alpha_{ij}$, with $i,j=0,1,\ldots$. Mutual exclusion 
statistics is manifested between the spaces $\cH_i$ and $\cH_j$ 
($i\ne j$) if $\alpha_{ij}\ne 0$--we say that the addition of 
$\delta N_j$ particles 
in the space $\cH_j$ changes the number of available states in the space 
$\cH_i$ by $-\alpha_{ij}\delta N_j$. 
With these notations, the total number of configurations is 
\cite{PhysRevLett.73.922.1994.Wu} 
\begin{equation} \label{conf_number1}
W = \prod_i\frac{\left[G_i+N_i-1-\sum_j\alpha_{ij}(N_j-\delta_{ij})
\right]!}
{N_i!\left[G_i-1-\sum_j\alpha_{ij}(N_j-\delta_{ij})\right]!} .
\end{equation}
Having the number of microscopic configurations (\ref{conf_number1}), 
if we asign the energy $\epsilon_i$ and the chemical potential $\mu_i$ 
to the states in the space $i$, we can calculate the  grandcanonical 
partition function, $\cZ$ \cite{PhysRevLett.73.922.1994.Wu}, 
\begin{equation}
\cZ = \sum_{\{N_i\}} W(\{N_i\})\exp\left[
\sum_i \beta N_i(\mu_i-\epsilon_i) \right]\,, \label{cZ_gen}
\end{equation}
and the total energy of the system in the given configuration, 
$E=\sum_i N_i\epsilon_i$--we use the notation 
$\beta=1/k_B T$, where $T$ is the temperature of the system. 

The most probable configuration, $\{N_i\}$, is obtained by 
maximizing $\cZ$ with respect to the set $\{N_i\}$. 
%For consistency with Ref. \cite{PhysRevLett.73.922.1994.Wu}, we used for the moment a different chemical potential, $\mu_i$, for each of the Hilbert spaces--since they have different exclusion statistics parameters, they could, for example, contain different species of particles. 
%
If we introduce the notations $n_i\equiv N_i/G_i$ and 
$\beta_{ij}\equiv\alpha_{ij}G_j/G_i$, and assume that for each $i$ 
both, $G_i$ and $N_i$ are sufficiently large, so that we can use the Stirling 
approximation [$\ln G_i! \approx G_i\ln(G_i/e)$ and 
$\ln N_i! \approx N_i\ln(N_i/e)$] the maximization procedure gives 
us the system of equations, 
\begin{equation}
n_i e^{\beta(\epsilon_i-\mu_i)} = 
\left[1+\sum_k(\delta_{ik}-\beta_{ik})n_k\right] 
\prod_j\left[\frac{1-\sum_k\beta_{jk}n_k} 
{1+\sum_k(\delta_{jk}-\beta_{jk})n_k}\right]^{\alpha_{ji}} \label{system_Wu}
\end{equation}
The system (\ref{system_Wu}) is solved easier if we denote 
$w_i\equiv n_i^{-1}-\sum_k\beta_{ik}n_k/n_i$. Using this notations 
(\ref{system_Wu}) becomes 
\begin{equation}
(1+w_i)\prod_j\left(\frac{w_j}{1+w_j}\right)^{\alpha_{ji}} = 
e^{\beta(\epsilon_i-\mu_i)} \label{EqforwWu}
\end{equation}
and $n_i$s can be calculated from the new system, 
\begin{equation}
\sum_j(\delta_{ij}w_j+\beta_{ij})n_j = 1 \,. \label{system_ni_Wu}
\end{equation}

If the spaces $i$ are fixed, 
Eq. (\ref{system_ni_Wu}) can be solved and the populations 
$n_i$ can be determined. 
But in general, the spaces $\cH_i$ can be changed by dividing or 
combining them and, if the system has a proper thermodynamic 
behavior, such changes should not affect the thermodynamic results. 
Unfortunately this is not the case and I will show now by simple 
examples that the particle distribution that maximizes the partition 
function depends on our choice of subspaces $\cH_i$. 

Let us assume that we have only two finite 
dimensional Hilbert spaces, $\cH_0$ and $\cH_1$, of dimensions $G_0$ and 
$G_1$, and the FES parameters $\alpha_{00}$, $\alpha_{11}$, $\alpha_{01}$, and 
$\alpha_{10}$. Then Eqs. (\ref{system_Wu}) and (\ref{EqforwWu}) reduce to 
systems of two equations and two unknowns: 
\begin{eqnarray}
\left\{\begin{array}{l}
(1+w_0)\left(\frac{w_0}{1+w_0}\right)^{\alpha_{00}}
\left(\frac{w_1}{1+w_1}\right)^{\alpha_{10}} = e^{\beta(\epsilon_0-\mu_0)} \\ \\
(1+w_1)\left(\frac{w_0}{1+w_0}\right)^{\alpha_{01}}
\left(\frac{w_1}{1+w_1}\right)^{\alpha_{11}} = e^{\beta(\epsilon_1-\mu_1)} 
\end{array} \right. \label{system11}
\end{eqnarray}
and 
\begin{eqnarray}
\left\{\begin{array}{l}
(w_0+\alpha_{00})n_0 + \alpha_{01}\frac{G_1}{G_0}n_1 = 1 \\ \\
\alpha_{10}\frac{G_0}{G_1}n_0 + (w_1+\alpha_{11})n_1 = 1 
\end{array}\right. \label{system21}
\end{eqnarray}
respectively. Now let us further assume that the space $\cH_1$ was 
obtained as a union of the two smaller, disjoint subspaces, 
$\cH_{1a}$ and $\cH_{1b}$, of dimensions $G_{1a}$ and $G_{1b}$, 
respectively. Then, instead of two Hilbert spaces and four 
exclusion statistics parameters that we started with, we can also describe the 
system as consisting of three subspaces and nine statistics parameters. 
From these nine, $\alpha_{00}$ remains unchanged, whereas 
$\alpha_{01a}$ and $\alpha_{01b}$ should be identical to $\alpha_{01}$. 
A natural choice for the other 
six parameters is $\alpha_{1a1a}=\alpha_{1b1b}\equiv\alpha_{11}$, 
$\alpha_{1b0}=\alpha_{1a0}=\alpha_{10}$, and 
$\alpha_{1a1b}=\alpha_{1b1a}=0$. Obviously, we also have 
$\epsilon_{1a}=\epsilon_{1b}=\epsilon_1$ and $\mu_{1a}=\mu_{1b}=\mu_1$. 
In the new configuration, the systems (\ref{EqforwWu}) and 
(\ref{system_ni_Wu}) become 
\begin{eqnarray}
\left\{\begin{array}{l}
(1+w'_0)\left(\frac{w'_0}{1+w'_0}\right)^{\alpha_{00}}
\left(\frac{w_{1a}}{1+w_{1a}}\right)^{\alpha_{10}}\left(\frac{w_{1b}}{1+w_{1b}}
\right)^{\alpha_{10}} = e^{\beta(\epsilon_0-\mu_0)} \\ \\
(1+w_{1a})\left(\frac{w'_0}{1+w'_0}\right)^{\alpha_{01}}
\left(\frac{w_{1a}}{1+w_{1a}}\right)^{\alpha_{11}} = e^{\beta(\epsilon_1-\mu_1)} 
\\ \\
(1+w_{1b})\left(\frac{w'_0}{1+w'_0}\right)^{\alpha_{01}}
\left(\frac{w_{1b}}{1+w_{1b}}\right)^{\alpha_{11}} = e^{\beta(\epsilon_1-\mu_1)} 
\end{array} \right. \label{system12}
\end{eqnarray}
and 
\begin{eqnarray}
\left\{\begin{array}{l}
(w'_0+\alpha_{00})n_0 + \alpha_{01}\frac{G_{1a}}{G_0}n_{1a} + 
\alpha_{01}\frac{G_{1b}}{G_0}n_{1b} = 1 \\ \\
\alpha_{10}\frac{G_0}{G_{1a}}n_0 + (w_{1a}+\alpha_{11})n_{1a} = 1  \\ \\
\alpha_{10}\frac{G_0}{G_{1b}}n_0 + (w_{1b}+\alpha_{11})n_{1b} = 1 
\end{array}\right. \label{system22}
\end{eqnarray}
The systems (\ref{system11}) and (\ref{system21}) should admit the 
same physical solution as the systems (\ref{system12}) and (\ref{system22}), 
which is $n_{1a}=n_{1b}=n_1$. If we also take, for the simplicity of the 
calculations, $G_{1a}=G_{1b}=G_1/2$, then from the last two equations of 
(\ref{system22}) we get $w_{1a}=w_{1b}$ and 
\begin{equation}
\alpha_{10}n_0G_0/G_1 + (w_{1a}-w_1)n_1 = 0 . \label{eqw1aw1}
\end{equation}
Using $w_{1a}=w_{1b}\equiv w_1'$ into (\ref{system12}), we obtain the system 
\begin{eqnarray}
\left\{\begin{array}{l}
(1+w'_0)\left(\frac{w'_0}{1+w'_0}\right)^{\alpha_{00}}
\left(\frac{w_{1}'}{1+w_{1}'}\right)^{2\alpha_{10}} 
= e^{\beta(\epsilon_0-\mu_0)} \\ \\
(1+w_{1}')\left(\frac{w'_0}{1+w'_0}\right)^{\alpha_{01}}
\left(\frac{w_{1}'}{1+w_{1}'}\right)^{\alpha_{11}} = e^{\beta(\epsilon_1-\mu_1)} 
\end{array} \right. \label{system13}
\end{eqnarray}
Since the systems (\ref{system11}) and (\ref{system13}) are not identical, 
Eqs. (\ref{system13}) and (\ref{eqw1aw1}) (eventually combined also 
with \ref{system11}) give three equations for the two unknowns, $w_0'$ and 
$w_1'$. The system of equations is overdetermined and does not have 
solutions, unless the statistics parameters have some special values. Such 
special values are e.g. $\alpha_{10}=\alpha_{01}=0$ 
($\alpha_{01}=0$ would be necessary if we would split the space $\cH_0$ instead 
of $\cH_1$). 

%But this procedure cannot be applied to our interacting system, because $n_i$s depend on our (arbitrar) choice of intervals. To see this, notice that the system (\ref{EqforwWu}) determines the quantities $w_i$ univoquely. If we plug $w_i$s into Eqs. (\ref{system_ni_Wu}), we notice that the solutions for $n_i$s depend on the dimensions of the intervals we choose through $\beta_{ij}=\alpha_{ij}G_j/G_i$. 

\subsection{Large systems}

I will show now that the thermodynamic inconsistency of 
FES happends also in systems of infinite, quasi-continuous 
spectra. For this let us assign to the 
single-particle states the sets of quantum numbers, $\bk$--the set $\bk$ 
does not necessary consist of wave-vector components, as it usually does, 
but it may contain any quantum numbers. If the system is large enough, we 
say that each of the subspaces $\cH_i$ contains the set of single particle 
states $\{\bk\}_i$, which ``fill-up'' a volume $V_i$ in the space of 
the quantum numbers and we define the density of 
states, $\sigma_\bk$, so that $G_i=\int_{V_i}\sigma_\bk\,d\bk$ 
and $N_i = \int_{V_i}n_\bk\sigma_\bk\,d\bk$. 
Then the parameters $\alpha_{ij}$ depend on the quantum 
numbers $\{\bk\}_i$ and $\{\bk\}_j$, so we shall change the subscripts and 
write in general, e.g. $\alpha_{\bk\bk'}$.
In these new notations, the system (\ref{EqforwWu}) becomes 
\begin{equation}
(1+w_\bk)\prod_{\bk'}\left(\frac{w_{\bk'}}{1+w_{\bk'}}\right)^{\alpha_{\bk'\bk}} = 
e^{\beta(\epsilon_\bk-\mu_\bk)} \label{EqforwWubk}
\end{equation}
for any division of the total Hilbert space of the system into elementary 
volumes $V_i$, as long as $G_i$ and $N_i$ (for any $i$) are large enough, 
so that we can apply the Stirling formula and the maximization procedure 
presented above. Moreover, if we can choose all the volumes $V_i$ so that 
they are small enough to use the approximation $G_i=\sigma_{\bk}V_i$ and 
$\alpha_{ij}=\alpha_{\bk\bk'}$ for any $\bk$ in $V_i$ and 
$\bk'$ in $V_j$, Eq. (\ref{system_ni_Wu}) becomes 
\begin{equation}
\sum_{\bk'}\left[\delta_{\bk\bk'}w_{\bk'}+\alpha_{\bk\bk'}\frac{\sigma_{\bk'}}
{\sigma_{\bk}}\cdot\frac{V_{\bk'}}{V_\bk}\right]n_{\bk'} = 1 \,, 
\label{system_ni_Wubk}
\end{equation}
where we also changed the notation $V_i$ into $V_\bk$ and 
$V_j$ into $V_{\bk'}$. 
Now we observe directly that while Eqs. (\ref{EqforwWubk}) contain 
only intensive parameters (i.e. which do not depend on the volumes 
$V_\bk$ involved), the values of the equilibrium particle populations 
(\ref{system_ni_Wubk}) depend on our arbitrary choice of volumes, so the 
thermodynamic quantities, which are the populations $n_\bk$, are 
not well defined. 

\section{Correction of the parameters} 

The reason for which the thermodynamics of FES systems is ambiguous is 
obvious from the beginning. If we look for example at Eq. (\ref{system_Wu}) 
and we imagine that we reduce the dimension of one of the spaces, say 
of space $i$, by half, than since all the parameters $\beta_{ij}$ are 
proportional to $G_j/G_i$, this could cause a significant 
resistribution of all the occupation numbers (as we proved in concrete cases 
above). Since the population numbers should be intensive parameters, it means 
that the system could not have thermodynamic limit. 

Fortunately, as it was easy to observe the anomaly of FES, it is as easy 
to correct it. I shall prove below that if the ``trouble-making'' mutual 
exclusion statistics parameters are proportional to the space on which 
they act, i.e. $\alpha_{ij}\propto G_i$, then the system admits a 
thermodynamic limit.
In Ref. \cite{submitted.FESinteraction} I analized the exclusion statistics 
that appears in general interacting systems and I showed that in the 
limit of a quasi-continuous density of states the mutual 
exclusion statistics parameters indeed satisfy this property. 
To differentiate these new parameters 
from the ones used above, I shall denote them by $\tilde\alpha_{\bk\bk'}$. 
Therefore, the qualitatively new feature of the parameters 
$\tilde\alpha_{\bk\bk'}$ 
is that for $\bk\ne\bk'$, $\tilde\alpha_{\bk\bk'}$ is not a simple number, 
but \textit{it is proportional to the dimension of the subspace} $V_\bk$: 
$\tilde\alpha_{\bk\bk'}=G_{V_\bk}\alpha_{\bk\bk'}\equiv\sigma_\bk V_\bk\alpha_{\bk\bk'}$. 
If $\bk=\bk'$, the statistics parameters are similar to the ones studied 
so far and we write $\tilde\alpha_{\bk\bk}=\alpha_{\bk\bk}$. 

Let us now try to find the equilibrium configuration with these 
new statistics parameters and see if this has a thermodynamic limit. 
To do this, we first have to write down the 
new number of configurations (\ref{conf_number1}):
%
%\begin{widetext}
\begin{equation} \label{conf_number_ext}
W = \prod_{\bk}\frac{\left[G_{V_\bk}+(1-\alpha_{\bk\bk})
(N_{V_\bk}-1)-G_{V_\bk}\sum^{\bk'\ne\bk}_{\bk'} 
\alpha_{\bk\bk'}N_{V_{\bk'}}\right]!}
{N_i!\left[G_{V_\bk}-1-\alpha_{\bk\bk}
(N_{V_\bk}-1)-G_{V_\bk}\sum^{\bk'\ne\bk}_{\bk'} 
\alpha_{\bk\bk'}N_{V_{\bk'}}\right]!} .
\end{equation}
%\end{widetext}
%
But we notice already that for large enough volumes $V_\bk$, both, 
$G_{V_\bk}$ and $N_{V_\bk}$ are large, therefore the summations 
$G_{V_\bk}\sum^{\bk'\ne\bk}_{\bk'}\alpha_{\bk\bk'}N_{V_{\bk'}}$, 
containing terms bilinear in these large quantities, 
must be much larger than $G_{V_\bk}$ and $N_{V_\bk}$. In conclusion both square 
brackets in (\ref{conf_number_ext}) should be negative. 
This is not admissible. [We note in passing 
that also in Haldane-Wu formalism (\ref{conf_number1}) 
\cite{PhysRevLett.67.937.1991.Haldane,PhysRevLett.73.922.1994.Wu} negative 
terms may appear in the calculation of the number of configurations 
for unappropriate choice of subspaces, even though in the summations 
there are only terms linear in $N_i$.] 

To solve the problem we have to regard the exclusion statistics as 
acting on the quasiparticle added as a perturbation to the equilibrium 
quasiparticle distribution. Therefore we assume that on top of the 
equilibrium distribution of particles, say $N_{V_\bk}$, we add a 
small perturbation, $\delta N_{V_\bk}$, which changes the number of 
configurations into 
\begin{eqnarray}
W &=& \prod_{\bk} \frac{[\tilde G_{V_\bk}+N_{V_\bk}+\delta N_{V_\bk}
-1-\sum^{\bk'\ne\bk}_{\bk'}\tilde\alpha_{\bk\bk'}\delta N_{V_{\bk'}}]!}
{(N_{V_\bk}+\delta N_{V_\bk})!(\tilde G_{V_\bk}-1-\sum^{\bk'\ne\bk}_{\bk'}
\tilde\alpha_{\bk\bk'}\delta N_{V_{\bk'}})!} \label{W_ext_pert}\\
&=& \prod_{\bk}\frac{\left[\tilde G_{V_\bk}+N_{V_\bk}+(1-\alpha_{\bk\bk})
\delta N_{V_\bk}-1-\tilde G_{V_\bk}\sum^{\bk'\ne\bk}_{\bk'}\alpha_{\bk\bk'} 
\delta N_{V_{\bk'}}\right]!}{(N_{V_\bk}+\delta N_{V_\bk})!
\left(\tilde G_{V_\bk}-\alpha_{\bk\bk}\delta N_{V_\bk}-1-\tilde G_{V_\bk} 
\sum^{\bk'\ne\bk}_{\bk'}\alpha_{\bk\bk'}\delta N_{V_{\bk'}}
\right)!} . \nonumber 
\end{eqnarray}
Notice that now $\tilde G_{V_\bk}$ is the Bose dimension of the Hilbert space, 
i.e. the number of available states, and not its real dimension. So, unless we 
have an ideal Bose system, the ratio $N_{V_\bk}/\tilde G_{V_\bk}$, denoted 
by $\tilde n_\bk$, is actually the 
ratio of the number of particles to the number of available states 
(or holes). In terms of $G_\bk$ of equation 
(\ref{conf_number1}) or (\ref{conf_number_ext}), 
$\tilde G_{V_\bk}=G_\bk-\alpha N_\bk$.
The equilibrium 
distribution is then obtained by imposing that the partition function is 
stationary with respect to the perturbations $\delta N_{V_\bk}$. 
If we introduce the notations, 
$\delta\tilde n_\bk\equiv\delta N_{V_\bk}/\tilde G_{V_\bk}$, using the Stirling 
approximation we write the logarithm of $W$ as 
\begin{eqnarray}
\fl
\ln W &=& \sum_\bk \tilde G_{V_\bk}\left\{
\left[1+\tilde n_\bk+(1-\alpha_{\bk\bk})\delta\tilde n_\bk-\sum_{\bk'(\ne\bk)}\alpha_{\bk\bk'}
\tilde G_{V_{\bk'}}\delta\tilde n_{\bk'}\right] \right. \nonumber \\
\fl
&& \times\ln\left[1+\tilde n_\bk+(1-\alpha_{\bk\bk})\delta\tilde n_\bk-\sum_{\bk'(\ne\bk)}
\alpha_{\bk\bk'}\tilde G_{V_{\bk'}}\delta\tilde n_{\bk'}\right]  \nonumber \\
\fl
&& -(\tilde n_\bk+\delta\tilde n_\bk)\ln(\tilde n_\bk+\delta\tilde n_\bk) - 
\left(1-\alpha_{\bk\bk}\delta\tilde n_\bk-1-\sum_{\bk'(\ne\bk)}\alpha_{\bk\bk'}
\tilde G_{V_{\bk'}}\delta\tilde n_{\bk'}\right) \nonumber \\ 
\fl
&&\left.\times\ln\left(1-\alpha_{\bk\bk}\delta\tilde n_\bk-1 
-\sum_{\bk'(\ne\bk)}\alpha_{\bk\bk'}\tilde G_{\bk'}\delta\tilde n_{\bk'}\right) \right\} 
 \label{lnW_ext_st2}
\end{eqnarray}
Adding $\beta\sum_\bk\tilde G_{V_\bk}\tilde n_\bk(\mu_\bk-\epsilon_\bk)$ to 
(\ref{lnW_ext_st2}), 
we obtain the logarithm of the partition function, which, if maximized, 
gives 
\begin{eqnarray}
\fl
0 &=& \frac{\partial}{\partial n_\bk}\left[
\ln W + \beta\sum_{\bk'}\tilde G_{V_{\bk'}}\tilde n_{\bk'}(\mu_{\bk'}-\epsilon_{\bk'})\right] 
\nonumber \\
\fl
&=& \tilde G_{V_\bk}\left\{ \beta(\mu_\bk-\epsilon_\bk) - \ln\tilde n_\bk + 
(1-\alpha_{\bk\bk})\ln(1+\tilde n_\bk) - \sum_{\bk'(\ne\bk)}\alpha_{\bk'\bk}\tilde G_{V_{\bk'}}\ln(1+\tilde n_{\bk'})\right\} \label{der_lnW_ext_st_gen}
\end{eqnarray}
We observe now that if $\alpha_{\bk\bk'}=0$, for any $\bk$ and $\bk'$, we 
obtain the Bose distribution, 
$\tilde n_\bk=n_\bk=\{\exp[\beta(\epsilon_\bk-\mu_\bk)-1]\}^{-1}$, whereas if 
$\alpha_{\bk\bk'}=0$, for any $\bk\ne\bk'$, and $\alpha_{\bk\bk}=1$ for any 
$\bk$, we obtain $\tilde n_\bk=\exp[\beta(\mu_\bk-\epsilon_\bk)]$, which is the 
ratio of the number of particles to the number of holes in a 
Fermi system, as expected. 

Now we can finally extract from (\ref{der_lnW_ext_st_gen}) a 
self-consistent equation for $n_\bk$: 
\begin{equation}
\beta(\mu_\bk-\epsilon_\bk)+\ln\frac{[1+\tilde n_\bk]^{1-\alpha_{\bk\bk}}}
{\tilde n_\bk} = \sum_{\bk'(\ne\bk)} G_{V_{\bk'}}\ln[1+\tilde n_{\bk'}] 
\alpha_{\bk'\bk} \label{inteq_for_n1}
\end{equation}
In the thermodynamic limit, when $G_{V_\bk}\ll\sum_{\bk'(\ne\bk)} G_{V_{\bk'}}$, 
Eq. (\ref{inteq_for_n1}) gives a good, unambiguous particle distribution 
in the system. The summation to the right-hand side of Eq. 
(\ref{inteq_for_n1}) can be readily and consistently transformed into 
an intergral when the dimension of the system increases and we get the 
self-consistent integral equation for the particle distribution, 
\begin{equation}
\beta(\mu_\bk-\epsilon_\bk)+\ln\frac{[1+\tilde n_\bk]^{1-\alpha_{\bk\bk}}}
{\tilde n_\bk}=\int \sigma_{\bk'}\ln[1+\tilde n_{\bk'}]\alpha_{\bk'\bk}\,d\bk' .
\label{inteq_for_n2}
\end{equation}

In the end let us remark an identity. If all the mutual 
statistics parameters were zero, the model introduced here should be 
identical to the previous FES model (section \ref{inconsistent}) so  
so let us compare the results. 
First, if $\alpha_{\bk\bk'}=0$ for all $\bk\ne\bk'$, equation 
(\ref{inteq_for_n1}) reduces to 
\begin{equation}
\frac{(1+\tilde n_\bk)^{1-\alpha_{\bk\bk}}}{\tilde n_\bk} = 
\e^{\beta(\epsilon_\bk-\mu_\bk)}, \label{inteq_for_n3}
\end{equation}
whereas equation (\ref{EqforwWubk}) becomes 
\begin{equation}
\frac{w_{\bk}^{\alpha_{\bk\bk}}}{(1+w_{\bk})^{\alpha_{\bk\bk}-1}} = 
\e^{\beta(\epsilon_\bk-\mu_\bk)} \label{EqforwWubk1}
\end{equation}
Now one can check easily that equation (\ref{inteq_for_n3}) is 
identical to equation (\ref{EqforwWubk1}) if $\tilde n_\bk=w_\bk^{-1}$.
But on the other hand, by definition, 
$w_\bk^{-1}=N_\bk/(G_\bk-\alpha_{\bk\bk}N_\bk)=N_\bk/\tilde G_\bk\equiv\tilde n_\bk$. Therefore the results are indeed identical and we observe with this ocasion 
that in FES systems with no mutual statistics, the quantity 
$w_\bk$ was the ration between the Bose dimension of the subspace $V_\bk$ 
and the number of particles in this subspace. 

\section{Conclusions}

I showed that fractional exclusion statistics leads to 
ambiguous thermodynamic results if the mutual exclusion statistics 
parameters are not zero ($\alpha_{ij}\ne 0$, for $i\ne j$). To correct 
this ambigiuty and to obtain consistent thermodynamic results I introduced 
new mutual exclusion statistics parameters, which are proportional to the 
dimension of the subspace on which they act. 

In a related publication \cite{submitted.FESinteraction} I proved that 
a gas of interacting particles can be described as a fractional 
exclusion statistics gas, with the exclusion statistics parameters 
having the properties obtained here. 

%\section{Acknowledgements}
\ack

I thank Dr. A. P\^arvan for motivating discussions. This work 
was done at the Bogoliubov Laboratory of Theoretical Physics, 
JINR Dubna, Russia and I thank the stuff of the 
Laboratory, especially Dr. S. N. Ershov and Dr. A. P\^arvan, for 
hospitality. The work was partially supported by the NATO grant, 
EAP.RIG 982080. 

\section*{References}
%\bibliography{/home/dragos/general}

\begin{thebibliography}{10}

\bibitem{PhysRevLett.67.937.1991.Haldane}
F.~D.~M. Haldane.
\newblock {\em Phys. Rev. Lett.}, 67:937, 1991.

\bibitem{ProgrTheorPhys.5.544.1950.Tomonaga}
S.~Tomonaga.
\newblock {\em Progr. Theor. Phys.}, 5:544, 1950.

\bibitem{JMathPhys.4.1154.1963.Luttinger}
J.~M. Luttinger.
\newblock {\em J. Math. Phys.}, 4:1154, 1963.

\bibitem{JMathPhys.6.304.1965.Mattis}
D.~C. Mattis and E~Lieb.
\newblock {\em J. Math. Phys.}, 6:304, 1965.

\bibitem{PhysRevLett.81.489.1998.Carmelo}
J.~M.~P. Carmelo, P.~Horsch, A.~A. Ovchinnikov, D.~K. Campbell, A.~H.
  Castro~Neto, and N.~M.~R. Peres.
\newblock {\em Phys. Rev. Lett.}, 81:489, 1998.

\bibitem{JMathPhys.10.2191.1969.Colagero}
F.~Colagero.
\newblock {\em J. Math. Phys. (N.Y.)}, 10:2191, 1969.

\bibitem{JMathPhys.12.247.1971.Sutherland}
B.~Sutherland.
\newblock {\em J. Math. Phys. (N.Y.)}, 12:247, 1971.

\bibitem{PhysRevA.4.2019.1971.Sutherland}
B.~Sutherland.
\newblock {\em Phys. Rev. A}, 4:2019, 1971.

\bibitem{PhysRevA.5.1372.1972.Sutherland}
B.~Sutherland.
\newblock {\em Phys. Rev. A}, 5:1372, 1972.

\bibitem{PhysRevB.60.6517.1999.Murthy}
M.~V.~N. Murthy and R.~Shankar.
\newblock {\em Phys. Rev. B}, 60:6517, 1999.

\bibitem{PhysRevLett.72.600.1994.Veigy}
A.~D. de~Veigy and S.~Ouvry.
\newblock {\em Phys. Rev. Lett.}, 72:600, 1994.

\bibitem{NuclPhysB470.291.1996.Hansson}
T.H. Hansson, J.M. Leinaas, and S.~Viefers.
\newblock {\em Nucl. Phys. B}, 470:291, 1996.

\bibitem{IntJModPhysA12.1895.1997.Isakov}
S.B. Isakov and S.~Viefers.
\newblock {\em Int. J. Mod. Phys. A}, 12:1895, 1997.

\bibitem{PhysRevLett.73.3331.1994.Murthy}
M.~V.~N. Murthy and R.~Shankar.
\newblock {\em Phys. Rev. Lett.}, 73:3331, 1994.

\bibitem{PhysRevLett.74.3912.1995.Sen}
D.~Sen and R.~K. Bhaduri.
\newblock {\em Phys. Rev. Lett.}, 74:3912, 1995.

\bibitem{JPhysB33.3895.2000.Bhaduri}
R.~K. Bhaduri, S.~M. Reimann, S.~Viefers, A.~G. Choudhury, and M.~K.
  Srivastava.
\newblock {\em J. Phys. B}, 33:3895--3903, 2000.

\bibitem{PhysRevLett.86.2930.2001.Hansson}
T.~H. Hansson, J.~M. Leinaas, and S.~Viefers.
\newblock {\em Phys. Rev. Lett.}, 86:2930--2933, 2001.

\bibitem{PhysRevLett.73.2150.1994.Isakov}
S.~B. Isakov.
\newblock {\em Phys. Rev. Lett.}, 73(16):2150, 1994.

\bibitem{PhysRevLett.73.922.1994.Wu}
Yong-Shi Wu.
\newblock {\em Phys. Rev. Lett.}, 73:922, 1994.

\bibitem{PhysRevLett.80.1698.1998.Iguchi}
K.~Iguchi.
\newblock {\em Phys. Rev. Lett.}, 80:1698, 1998.

\bibitem{PhysRevB.61.12757.2000.Iguchi}
K.~Iguchi.
\newblock {\em Phys. Rev. B}, 61:12757, 2000.

\bibitem{PhysRevLett.85.2781.2000.Iguchi}
K.~Iguchi and B.~Sutherland.
\newblock {\em Phys. Rev. Lett.}, 85:2781, 2000.

\bibitem{submitted.FESinteraction}
D.~V. Anghel.
\newblock arXiv:0710.0728, 2007.

\end{thebibliography}
%\bibliographystyle{unsrt}

\end{document}